\documentclass{JFM-FLM_Au}

\usepackage{orcidlink}

\lefttitle{Z. Bramantasaputra, D. D. Wangsawijaya and B. Ganapathisubramani}
\righttitle{Journal of Fluid Mechanics}

\title{Effects of spatially localised pressure gradient histories on recovery of turbulent boundary layers}

\author{Zefanya Bramantasaputra\aff{1}\orcidlink{0009-0001-2878-8085}, Dea Daniella Wangsawijaya\aff{1}\orcidlink{0000-0002-7072-4245
} \and Bharathram Ganapathisubramani\aff{1}\orcidlink{0000-0001-9817-0486}}

\affiliation{\aff{1}Department of Aeronautical and Astronautical Engineering, University of Southampton, Southampton SO16 7QF, United Kingdom}

\corresau{Zefanya Bramantasaputra, \href{mailto:zb2n23@soton.ac.uk}zb2n23@soton.ac.uk}

\usepackage{enumitem}
\usepackage{pdflscape}
\usepackage{overpic}
\hypersetup{hidelinks}

\usepackage{tikz}
\usetikzlibrary{shapes.geometric,shapes.symbols}
\tikzset{symbol style/.style={ultra thick, fill=none, minimum size=3mm, inner sep=0pt}}

\definecolor{b}{RGB}{30.6,119.85,181.05}
\definecolor{g}{RGB}{43.35,160.65,43.35}
\definecolor{o}{RGB}{255,127.5,12.75}

\definecolor{b1}{RGB}{7.65,48.45,107.1}
\definecolor{b2}{RGB}{7.65,79.05,153}
\definecolor{b3}{RGB}{30.6,109.65,178.5}
\definecolor{b4}{RGB}{61.2,140.25,193.8}
\definecolor{b5}{RGB}{96.9,165.75,209.1}
\definecolor{b6}{RGB}{142.8,193.8,221.85}
\definecolor{b7}{RGB}{183.6,214.2,234.6}

\definecolor{g1}{RGB}{0,68.85,28.05}
\definecolor{g2}{RGB}{0,107.1,43.35}
\definecolor{g3}{RGB}{30.6,135.15,66.3}
\definecolor{g4}{RGB}{61.2,165.75,89.25}
\definecolor{g5}{RGB}{104.55,191.25,112.2}
\definecolor{g6}{RGB}{147.9,211.65,142.8}
\definecolor{g7}{RGB}{186.15,226.95,178.5}

\definecolor{o1}{RGB}{127.5,38.25,5.1}
\definecolor{o2}{RGB}{163.2,53.55,2.55}
\definecolor{o3}{RGB}{211.65,68.85,0}
\definecolor{o4}{RGB}{237.15,99.45,15.3}
\definecolor{o5}{RGB}{249.9,132.6,51}
\definecolor{o6}{RGB}{252.45,163.2,91.8}

\definecolor{k1}{RGB}{0,0,0}
\definecolor{k2}{RGB}{102,102,102}
\definecolor{k3}{RGB}{204,204,204}

\begin{document}
\maketitle

\begin{abstract}
Hot-wire anemometry is used to investigate the recovery of smooth-wall turbulent boundary layers from spatially localised (i.e. impulsive) pressure gradient history (PGH) effects. Measurements are performed at multiple stations downstream of spatial distributions of favourable-adverse pressure gradient sequences, followed by relaxation to zero-pressure-gradient (ZPG) conditions. The analysis focuses on matched friction Reynolds numbers at $Re_\tau \approx 2300$, $3000$, and $5500$ (closest to farthest downstream of the imposed impulses), where the local Clauser pressure gradient (PG) parameter $\beta$ is nominally matched at $1.7$, $0.6$, and $-0.1$, respectively. PGH strength is quantified using the integral history parameter $\upDelta\beta$, proposed by \href{https://doi.org/10.1017/jfm.2025.320}{Preskett \textit{et al.} (\textit{J. Fluid Mech.}, vol. 1010, 2025, A30)}, which allows isolation of PGH as the primary source of variation. The imposed PGH amplifies the wake component of the mean velocity profile and enhances the streamwise Reynolds stress throughout the boundary layer, including the emergence of an outer peak. Spectral analysis reveals an additional outer-layer energetic feature with streamwise length scales of $2$--$3\delta$ ($\delta$ is the local boundary layer thickness), identified as the PG peak, distinguishable from the very-large-scale motion (VLSM). Even after $\beta$ has relaxed (to zero) for sufficiently long distances, mean flow has not recovered to ZPG state. Once $\upDelta\beta \lesssim 0.1$, mean flow and inner/log-layer turbulence statistics recover; however, the outer-layer turbulence retains a persistent imprint of PGH. Finally, we observe that recovery involves reorganisation of large-scale structures -- where VLSMs appear to be shorter even after the PG peak has vanished -- which indicates prolonged history effects.
\end{abstract}

\begin{keywords}
\end{keywords}


\section{\label{sec:Introduction}Introduction}

Turbulent boundary layers (TBLs) subjected to streamwise pressure gradients (PGs) are ubiquitous in engineering applications, including flows over aircraft wings, wind turbines, and ship hulls. The similarity between these applications, from a flow perspective, is that they eventually experience high Reynolds numbers while also being exposed to strong PGs early in their evolution. The imposition of local PGs on canonical zero-pressure-gradient (ZPG) smooth-wall TBLs alters not only the mean flow but also the underlying turbulence structures \citep{Lee2009, Harun2013}. A further level of complexity arises from \emph{history} effects: \citet{Clauser1956} noted that the outer portion of the boundary layer possesses a long memory. This means that the downstream state of a TBL depends not only on the \emph{local} PG condition but also on the upstream sequence of PG conditions, here referred to as pressure gradient history (PGH).

High-Reynolds-number canonical ZPG TBLs exhibit complex multiscale organisation, characterised by a hierarchy of coherent structures spanning from near-wall streaks to large-scale motions (LSMs) and very-large-scale motions (VLSMs). LSMs contribute significantly to Reynolds shear stress, with streamwise extents of 2--3$\delta$, where $\delta$ is the boundary layer thickness \citep{Townsend1951, Adrian2000, Ganapathisubramani2003}. Here, $\delta$ is taken as the wall-normal location at which $U = 0.99U_\infty$, where $U_\infty$ is the freestream velocity. VLSMs emerge at sufficiently high friction Reynolds numbers ($Re_\tau \gtrsim 2000$, with $Re_\tau \equiv U_\tau \delta / \nu$, where $U_\tau$ is the friction velocity and $\nu$ the kinematic viscosity) and become increasingly significant as $Re_\tau$ increases -- premultiplied energy spectra show that they typically occur at $y^+ \approx 3.9\sqrt{Re_\tau}$ (with $y^+ \equiv y U_\tau / \nu$, where $y$ is the wall-normal location), corresponding to $y/\delta \approx 0.06$, with characteristic streamwise length scales of $\lambda_x \approx 6\delta$ \citep{Kim1999, Hutchins2007, Mathis2009}. These structures modulate small-scales in the near-wall region, influencing velocity gradients and skin-friction drag \citep{Hutchins2011, Ganapathisubramani2012}. 

While complex in structure, canonical TBLs can be characterised by the classical logarithmic law of the wall, which describes the streamwise mean velocity profile in the overlap region, given by $U^+ = (1/\kappa) \ln (y^+) + B$, where $U^+ \equiv U/U_\tau$, and $\kappa$ and $B$ are the von K{\'a}rm{\'a}n constant and the log law intercept, respectively. For high-$Re_\tau$ canonical TBLs, analyses of available datasets have suggested $\kappa = 0.39 \pm 0.02$ \citep{Marusic2013}, while $B$ is often considered to be closely linked to $\kappa$ \citep{Nagib2008, Baxerres2024}. In the outer-layer (also referred to as the defect or wake region), the mean velocity profile can be extended following the law of the wake \citep{Coles1956}, such that
\begin{equation}
    U^+ = \frac{1}{\kappa} \ln (y^+) + B + \frac{2\mathit{\Pi}}{\kappa}\mathcal{W}(\eta),
\end{equation} 
where $\mathit{\Pi}$ is the Coles wake parameter and $\mathcal{W}(\eta)$ is the wake function. \citet{Chauhan2009} proposed an exponential-type formulation for $\mathcal{W}(\eta)$, yielding a typical value of $\mathit{\Pi} \approx 0.44$ for canonical TBLs; this formulation is adopted in the present study. It should be noted that the definition of the wake function is not universal, and different formulations may lead to variations in the reported values of $\mathit{\Pi}$.

The presence of streamwise PGs alters the mean flow: a favourable PG (FPG, $\mathrm{d}p/\mathrm{d}x<0$) accelerates the freestream, while an adverse PG (APG, $\mathrm{d}p/\mathrm{d}x>0$) has the opposite effect, where $p$ is the static pressure and $x$ the streamwise coordinate \citep{Schlichting2018, Narasimha1973, Simpson1989}. To characterise the influence of PGs on TBLs, \citet{Clauser1956} introduced a \emph{local} PG parameter, $\beta$, defined as $\beta = (\delta^* / \tau_w) (\mathrm{d}p / \mathrm{d}x)$, with $\delta^*$ the displacement thickness given by $\delta^* = \int_{0}^{\infty} \left( 1 - U/U_\infty \right) \mathrm{d}y$, where $U$ is the mean streamwise velocity at a given $y$, and $\tau_w$ is the wall shear stress. Although $\beta$ is widely adopted, its interpretation as a purely local parameter is not without limitations; for example, \citet{Maciel2018} showed that its physical meaning becomes less clear for flows with large shape factors ($H > 1.6$) and at very large Reynolds numbers. Nevertheless, maintaining a constant or near-constant $\beta$ -- typically achieved by imposing an approximately constant $\mathrm{d}p/\mathrm{d}x$ over the streamwise development -- has traditionally been regarded as a prerequisite for equilibrium or quasi-equilibrium conditions, enabling the effects of the local PG to be isolated \citep{Gungor2016, Devenport2022}. Since the present study focuses on attached TBLs (where $H$ remains close to canonical values) at high yet finite Reynolds numbers, $\beta$ is adopted here as a consistent measure of the local streamwise PG strength.

Studies of near-equilibrium PG TBLs have established the principal effects of \emph{local} PGs on both mean flow and turbulence statistics, as summarised in table \ref{Table:1} \citep{Spalart1993, Harun2013, Vinuesa2017b, Monty2011}.
\begin{table}
\begin{center}
\setlength{\tabcolsep}{14pt}
    \begin{tabular}{lccc}
    \textbf{Quantity} & \textbf{APG ($\beta>0$)} & \textbf{FPG ($\beta<0$)} \\
    \hline
    Shape factor, $H$ & Increases $\nearrow$ & Decreases $\searrow$ \\
    Skin-friction coefficient, $C_f$ & Decreases $\searrow$ & Increases $\nearrow$ \\
    Wake parameter, $\mathit{\Pi}$ & Increases $\nearrow$ & Decreases $\searrow$ \\
    Streamwise Reynolds stress, $\overline{u^2}$ & Increases $\nearrow$ & Decreases $\searrow$ \\
    \end{tabular}
    \caption{Effects of PGs on boundary layer quantities in near-equilibrium PG TBLs relative to ZPG TBL.}
    \label{Table:1}
\end{center}
\end{table}
Of particular relevance here, APGs elevate the streamwise Reynolds stress ($\overline{u^2}$, i.e. the variance of the streamwise velocity fluctuations) and promote the emergence of a pronounced outer peak, even at relatively low friction Reynolds numbers where such features are absent in canonical ZPG flows \citep{Monty2011, Kitsios2016, Deshpande2023}. This behaviour is associated with increased energy in large-scale motions within the outer-layer, reflecting enhanced turbulence production away from the wall \citep{Bobke2017, Gungor2022}. In contrast, the near-wall region remains comparatively robust, with the inner peak location and small-scale structures largely preserved despite significant outer-layer modification \citep{Harun2013, Pozuelo2022a}. It is worth noting that the presence of PGs may lead to deviations from classical scaling behaviour. In particular for APGs, the logarithmic region may be shortened or altered, and the universality of inner and outer scaling becomes less clear as $\beta$ increases \citep{Knopp2021, Deshpande2024}, reflecting the growing influence of outer-layer dynamics. Along with this, the effect of PGs on the logarithmic-law coefficients, $\kappa$ and $B$, remains an active topic of discussion: while some studies report these coefficients to remain approximately constant under mild APGs (e.g. $\beta < 2.3$) \citep{Aubertine2005, Volino2020}, others observe that although $\kappa$ is largely unaffected, the intercept $B$ varies with $\beta$ \citep{Monty2011, Zarei2026}. However, the accurate determination of these parameters is highly sensitive to uncertainties in skin-friction measurements, which can significantly influence inner normalisation. As the present study estimates skin friction using an indirect method (see \S~\ref{subsec:TBL properties}), while the range of $\beta$ considered lies within the mild APG regime, $\kappa$ and $B$ are treated as constants; consequently, any detailed assessment of their variation is beyond the scope of this work.

In the presence of PGH, local descriptors such as $\beta$ and its streamwise variation $\mathrm{d}\beta/\mathrm{d}x$ (i.e. the local disequilibrium effect) are insufficient to capture the physics of such non-equilibrium TBLs, as they do not account for the accumulated influence of upstream conditions \citep{Gungor2024}. Some efforts to account for PGH were based on the average value of $\beta$ up to a given Reynolds number based on momentum thickness, $\theta$, denoted as $\overline{\beta}(Re_\theta)$ \citep{Vinuesa2017a}. This framework has since been extended in various forms, including alternative averaging limits and reduced-order representations of the upstream history \citep{Mahajan2026, Gomez2025, Virgilio2025}. However, such formulations require knowledge of the full $\beta(Re_\theta)$, or equivalently $\beta(x)$, with sufficient streamwise resolution over the upstream history of interest, which is not available in most experimental studies. An alternative formulation was proposed by \citet{Preskett2025a}, who introduced $\upDelta \beta$ as an integrated measure to account for PGH,
\begin{equation}
    \upDelta \beta (x) = \left[\frac{\delta^*(x)}{\tau_w(x)}\right] \frac{1}{L} \int_{x-L}^{x} \left[\frac{\mathrm{d}}{\mathrm{d}x'}p(x') \right]\, w(x')\,\mathrm{d}x',
\end{equation}
where $L$ is the streamwise integration length upstream of a given location $x$, and $w$ is a linear weighting function (with $w = 1$ at $x' = x$ and $w = 0$ at $x' = x-L$) that emphasises the greater influence of regions closer to the given location. This formulation avoids the need for full flow-field measurements to obtain $\delta^*(x)$ and $\tau_w(x)$ by assuming that their local values already embed the accumulated history effects, while still explicitly accounting for PGH through the weighted integration of the upstream PG distribution; this makes the approach particularly well suited for experimental work. A common feature of such PGH parameters is their dependence on a characteristic integration length scale, here denoted by $L$. Although previous studies have suggested values of order \textit{O}$(10\delta)$ \citep{Vinuesa2017a}, no consensus has yet emerged, and the appropriate history response length scale remains an active area of research \citep{Gungor2024}. For instance, \citet{Gomez2025} adopted a fixed integration length of $12\delta$, while \citet{Preskett2025b} reported a value of about $6\delta$ for rough-wall flows. Alternatively, \citet{Virgilio2025} proposed a reduced-order formulation that incorporates the full upstream history through a weighted sigmoid function. In the present study, following \citet{Preskett2025a}, an integration length of approximately $L \approx 30\delta$ is employed in the definition of $\upDelta \beta$, with the aim of capturing a sufficiently extended upstream history while maintaining consistency with prior work.

\citet{Preskett2025a} showed that TBLs subjected to non-equilibrium PG sequences (i.e., streamwise-varying $\beta$) remain significantly modified in terms of skin friction and mean velocity even \emph{after} returning to nominally ZPG conditions $(\beta \approx 0)$. They further demonstrated that their PGH integral parameter $\upDelta \beta$ captures this behaviour through a clear relationship with deviations in the wake parameter, $\upDelta \mathit{\Pi} = \mathit{\Pi} - \mathit{\Pi}_{\mathrm{ZPG}}$. However, it is worth noting that characterising PGH effects solely through $\upDelta \mathit{\Pi}$ may be insufficient, as history effects can modify the extent of the overlap region -- manifested as a wall-normal stretching of the wake region -- which introduces uncertainty in the estimation of $\mathit{\Pi}$ \citep{Zarei2026}. In terms of turbulence statistics, \citet{Vila2017} and \citet{Tanarro2020} showed that PGH primarily affects the outer region, highlighting the dominant role of history effects over instantaneous local PG alone. More recently, \citet{Preskett2026} extended the use of $\upDelta \beta$ to turbulence energy measures, reporting an approximately linear scaling between $\upDelta \beta$ and the peak outer-region energy. Together, these observations emphasise the importance of accumulated \emph{history} effects and the need to distinguish them from \emph{local} PG effects, not only in mean flow quantities but also in turbulence statistics and coherent structures -- a challenge that remains central to PGH studies \citep{Vinuesa2017a, Bobke2017, Gungor2022}.

\begin{figure}
    \centering
    \begin{overpic}[width=\textwidth]{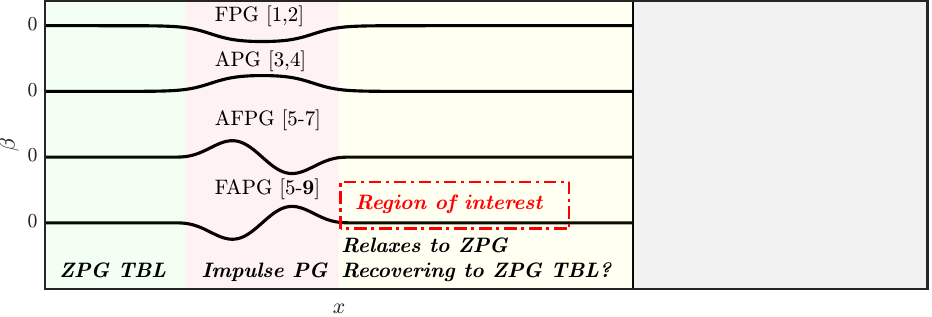}
        \put(69,32){\footnotesize [1] \citet{Volino2020}}
        \put(69,28.5){\footnotesize [2] \citet{Kumar2025}}
        \put(69,25){\footnotesize [3] \citet{Gungor2024}}
        \put(69,21.5){\footnotesize [4] \citet{Gomez2025}}
        \put(69,18){\footnotesize [5] \citet{Vishwanathan2023}}
        \put(69,14.5){\footnotesize [6] \citet{Virgilio2025}}
        \put(69,11){\footnotesize [7] \citet{Preskett2025a}}
        \put(69,7.5){\footnotesize [8] Parthasarathy \textit{et al.} \citeyearpar{Parthasarathy2023}}
        \put(69,4){\footnotesize \textbf{[9] Present study}}
    \end{overpic}
    \caption{Schematic representation of spatially impulsive PG forcing applied to a canonical ZPG TBL. The green region denotes the incoming canonical ZPG TBL, the red region indicates the imposed PG impulse, and the yellow region represents the downstream recovery region where the local PG parameter has relaxed to ZPG conditions $(\beta \approx 0)$. Four idealised impulse configurations are illustrated: FPG, APG, AFPG, and FAPG.}
    \label{Figure:1}
\end{figure}

One approach to isolating and investigating PGH effects is through the application of a spatially impulsive PG as illustrated schematically in figure \ref{Figure:1}. Here, the term \emph{impulse} refers to a localised PG forcing in the streamwise direction rather than a temporal one, with a finite streamwise extent that may range from relatively short to long compared to the incoming boundary layer thickness, $\delta_0$. Following an initial ZPG state, the flow is subjected to a finite region of imposed PG forcing before the local PG \emph{relaxes} back to nominally ZPG conditions $(\beta \approx 0)$. Depending on the imposed sequence, the impulsive PG may take the form of an FPG \citep{Volino2020, Kumar2025}, an APG \citep{Gungor2024, Gomez2025}, or combined sequences such as adverse-favourable PG (AFPG) and favourable-adverse PG (FAPG), representative of more complex aerodynamic flows such as those over aerofoils \citep{Vishwanathan2023, Virgilio2025, Preskett2025a, Parthasarathy2023}. While it may be expected that, sufficiently far downstream, the flow ultimately returns to a state governed solely by local conditions, it does not \emph{recover} immediately to a canonical state, even after $\beta \approx 0$ is reached \citep{Preskett2025a}. Here, \emph{relaxation} refers to the local PG returning to $\beta \approx 0$, whereas \emph{recovery} denotes the return of the boundary layer to a ZPG-like  state. Across these studies, attention has focused primarily on the imposed PG region itself (red region in figure \ref{Figure:1}), with comparatively little emphasis on the downstream recovery region (yellow region in figure \ref{Figure:1}); the only exception is the work of \citet{Volino2020}, who showed that the TBL recovers from an FPG impulse to a canonical ZPG-like state within approximately $40\delta_0$. However, for other forms of impulsive PG, the subsequent downstream recovery process remains unclear -- i.e. how does the TBL recover from a history-dominated state towards a locally governed equilibrium state?

Building on the findings of \citet{Preskett2025a} that \emph{recovery} does not happen simultaneously with \emph{relaxation}, this study focuses on the downstream recovery region of a mild FAPG impulse, following the experimental framework of \citet{Virgilio2025} and \citet{Preskett2025a}. While \citet{Virgilio2025} examined the flow \emph{undergoing} the impulsive PG sequence (i.e. primarily within the red region in figure \ref{Figure:1}), and \citet{Preskett2025a} focused on the flow immediately downstream of the imposed PG region, where the local PG has \emph{just} relaxed to $\beta \approx 0$ (i.e. at the onset of the yellow region in figure \ref{Figure:1}), the present study aims to address the following key question: how does a TBL subjected to an impulsive FAPG \emph{recover} further downstream after it \emph{relaxes}? Specifically, we aim to examine (i) the influence of PGH strength on the recovery of the mean flow and turbulence characteristics, and (ii) the evolution of the associated energetic scales during recovery.

To address these questions, wind tunnel experiments were performed on smooth-wall TBLs exposed to three distinct non-equilibrium FAPG histories, imposed by an aerofoil set at negative angles of attack, $\alpha$, which subsequently relaxes to nominally ZPG conditions. These sequences were designed with identical phase but varying amplitude, thereby isolating the influence of PGH strength; the configuration may be interpreted as an impulse-like FAPG excitation (i.e. a sudden FAPG perturbation) of varying magnitude. Hot-wire anemometry was employed in the downstream region to characterise both mean flow and turbulence characteristics, with measurements conducted at multiple streamwise locations where the flow approaches nominally ZPG conditions, enabling assessment of the recovery process. Comparisons are made at matched $Re_\tau$ and $\beta$, allowing an independent evaluation of history effects. It should be noted that the present study relies on an indirect estimation of $\tau_w$ (obtained via a fitting-based method, see \S~\ref{subsec:TBL properties}), which represents a limitation of the experimental approach.

The remainder of this paper is organised as follows. Section \ref{sec:Experimental setup and parameters} discusses the experimental methods used and the flow history experienced; we also present the measured TBL parameters here. Section \ref{sec:Matched Reynolds number} presents the mean flow and turbulence characteristics for selected cases at matched $Re_\tau$ and local $\beta$, isolating the effect of PGH strength. Downstream TBL measurements provided in \S~\ref{sec:Experimental setup and parameters} are further analysed, and their associated energetic scales are presented in \S~\ref{sec:Recovery observation} to examine the evolution of these scales during recovery. Conclusions are presented in \S~\ref{sec:Conclusion}.

\section{\label{sec:Experimental setup and parameters}Experimental setup and parameters}

\subsection{\label{subsec:Flow facility}Flow facility}
\begin{figure}
    \centering
    \includegraphics[width=\textwidth]{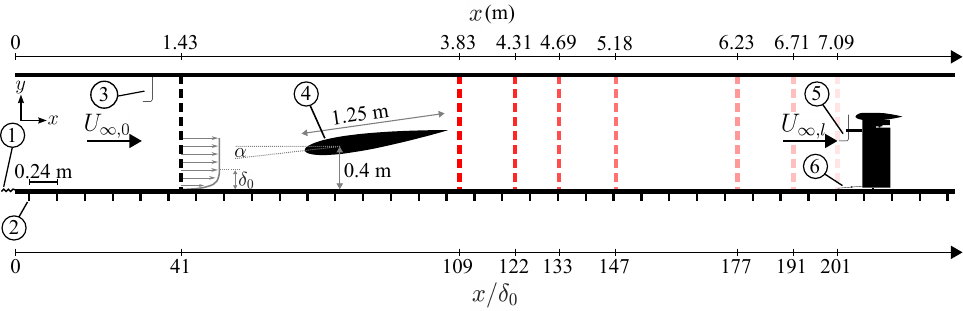}
    \caption{Schematic of the BLWT test section. The black dashed line indicates the upstream reference measurement location, where the reference boundary layer thickness upstream of the wing was measured as $\delta_0 = 35.23$ mm (not to scale in the schematic). The red dashed lines denote downstream measurement locations, with lighter shades corresponding to further downstream positions. All locations are normalised by $\delta_0$, corresponding to $x/\delta_0 =$ 41 (upstream), and 109, 122, 133, 147, 177, 191, and 201 (downstream). Numbered markers indicate key components: \protect\textcircled{1} boundary layer trip; \protect\textcircled{2} pressure tap array (starting 0.12 m from the test section start); \protect\textcircled{3} upstream Pitot tube used to set $U_{\infty,0}$; \protect\textcircled{4} NACA 0012 aerofoil generating the PG; \protect\textcircled{5} local Pitot tube measuring $U_{\infty,l}$ and serving as hot-wire calibration reference; \protect\textcircled{6} hot-wire probe.}
    \label{Figure:2}
\end{figure}

The experiments were conducted in the Boundary Layer Wind Tunnel (BLWT) at the University of Southampton. This closed-loop facility has a test section that consists of five segments, each 2.4 m long, giving a total streamwise length of 12 m, with a spanwise width of 1.2 m and wall-normal height of 1 m, where the streamwise, wall-normal, and spanwise directions are denoted by $(x, y, z)$, respectively. Before entering the test section, the airflow passes through a settling chamber equipped with a honeycomb and three screens to reduce turbulence, resulting in a measured freestream turbulence intensity of less than 0.15\%. A simplified schematic of the experimental setup is shown in figure \ref{Figure:2}, depicting most of the first four segments of the test section along with the key setup elements.

The floor is constructed from a polished flat plate, preceded by a smooth-wall ramp with an angle of $5^\circ$ to match the thickness of the test surface. Downstream of the ramp, a turbulator tape is attached to promote transition to turbulence, as indicated by marker \textcircled{1} in figure \ref{Figure:2}. 

The reference freestream velocity, $U_{\infty,0}$, and the local freestream velocity, $U_{\infty,l}$, were measured using the Pitot-static tubes at markers \textcircled{3} and \textcircled{5} in figure \ref{Figure:2}, each connected to a Furness Controls FCO560 manometer and recorded via an NI USB-6212 BNC DAQ module with 16-bit resolution. The velocity was determined from the measured dynamic pressure, with air density calculated under the perfect gas assumption using the ambient pressure and temperature recorded by the BLWT's embedded sensors, and viscosity obtained from Sutherland's law. In this study, $U_{\infty,0}$ was regulated to $20\pm0.5\%$ m/s using a PID controller, while the air temperature in the tunnel test section was maintained at $20\pm 2\% ^\circ$C by the BLWT's water-cooled heat exchanger system.

\subsection{\label{subsec:Flow history}Flow history}
The streamwise PG was generated using a 2D NACA0012 wing with a chord length of 1.25 m (see \textcircled{4} in figure \ref{Figure:2}) and a span of 118.5 cm, covering 99\% of the test section width. The wing was suspended from the test section roof using four linear actuators, allowing its attitude to be adjusted by extending or retracting the actuators. Its 25\% chord was positioned 280 cm from the start of the test section and set 40 cm above the floor. Three different flow histories, corresponding to varying PG magnitudes, were achieved by setting the wing angle of attack to $-4^\circ$, $-6^\circ$, and $-8^\circ$ while keeping the quarter-chord position fixed.

\begin{figure}
    \centering
    \includegraphics[width=\textwidth]{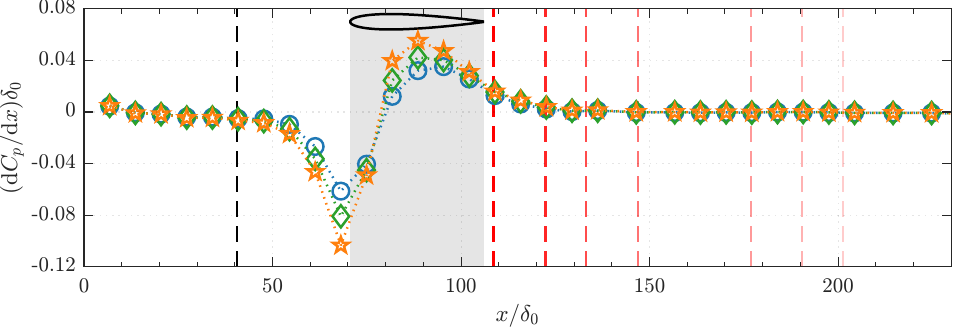}
    \caption{Streamwise pressure coefficient gradient $\mathrm{d}C_p/\mathrm{d}x$, normalised by the upstream boundary layer thickness $\delta_0$, for all three cases. 
    \protect\begin{tikzpicture}
        \protect\node[circle, draw=b, symbol style]{};\protect\draw[b, line width = 1.5pt, dash pattern=on 1pt off 1.5pt] (-0.3,0) -- (0.3,0);
    \protect\end{tikzpicture}, 
    \protect\begin{tikzpicture}
        \protect\draw[g, ultra thick, fill=none]
        (0,0.15) -- (0.12,0) -- (0,-0.15) -- (-0.12,0) -- cycle;\protect\draw[g, line width = 1.5pt, dash pattern=on 1pt off 1.5pt] (-0.3,0) -- (0.3,0);
    \protect\end{tikzpicture}, and
    \protect\begin{tikzpicture}
        \protect\node[star, star points=5, star point ratio=2.5, draw=o, symbol style]{};\protect\draw[o, line width = 1.5pt, dash pattern=on 1pt off 1.5pt] (-0.3,0) -- (0.3,0);\protect\end{tikzpicture} denote the symbols and colours used for the $-4^\circ$ (weak), $-6^\circ$ (mild), and $-8^\circ$ (strong) PGH cases, respectively. The grey shaded region indicates the presence of the wing, and vertical dashed lines mark the streamwise measurement locations described previously.}
    \label{Figure:3}
\end{figure}

To characterise the streamwise PG imposed on the TBL, the static pressure distribution along the test section floor was measured. Forty static pressure taps were installed across the first four segments of the smooth-wall test section, spaced 24 cm apart (see \textcircled{2} in figure \ref{Figure:2}). The taps were positioned 25 cm from the side wall (35 cm off-centre) to minimise disturbances to the flow in the central region where TBL measurements were performed. Pressures were recorded using a Scanivalve ZOC 33/64Px electronic pressure scanning module with a RAD3200 system. The pressure coefficient was calculated as
\begin{equation}
    C_p = \frac{{p-p_{0}}}{{0.5\rho U_{\infty,0}^2}},
\end{equation} 
where $p$ is the static pressure measured by the wall taps, $p_0$ is the reference static pressure at the tap nearest the Pitot tube measuring $U_{\infty,0}$ (fifth tap, just below \textcircled{3} in figure \ref{Figure:2}), and $\rho$ is the fluid density. The normalised streamwise PG coefficient, $(\mathrm{d}C_p/\mathrm{d}x)\delta_0$, for the three cases investigated in this study is shown in figure \ref{Figure:3}.

Upstream of the wing, the flow is nominally ZPG, with a slight FPG due to boundary layer growth, extending to the reference point at $x/\delta_0 = 41$. Exposure to the wing generates a characteristic FAPG impulse, with amplitude increasing as $\alpha$ decreases from $-4^\circ$ to $-6^\circ$ and $-8^\circ$. In all cases, the impulse exhibits a similar FAPG phase distribution, with approximately $40\delta_0$ of FPG followed by $50\delta_0$ of APG, after which the flow returns to ZPG conditions approximately one chord length downstream of the trailing edge. Despite a similar setup to \citet{Preskett2025a}, the current impulsive PG develops over roughly $90\delta_0$, about twice the lengthscale of the previous study, due to the wing being relocated 4.04 m upstream. Across all cases, variations in impulse amplitude far exceed local PG differences at the same measurement locations, so any downstream TBL differences are attributed to their respective upstream histories. For example, at the first downstream measurement station ($x/\delta_0 = 109$), the difference in upstream APG history between the $\alpha = -4^\circ$ and $-8^\circ$ cases (evaluated around $x/\delta_0 \approx 90$) is approximately a factor of 7 larger than the corresponding local PG difference. Accordingly, the cases are hereafter referred to as PGH cases, with $\alpha = -4^\circ$, $-6^\circ$, and $-8^\circ$ corresponding to weak, mild, and strong PGH, respectively. It is emphasised that this classification (weak, mild, strong PGH) is specific to the present PGH study framework and should not be confused with classifications based on instantaneous APG strength. In terms of local PG parameter conditions, all downstream measurement locations where $\beta$ is \emph{not} nominally zero lie within the mild APG regime ($\beta < 2.3$, see table \ref{Table:2}). Note that the PG data were measured with the hot-wire traverse system removed.

\subsection{\label{subsec:Hot-wire anemometry}Hot-wire anemometry}
Hot-wire anemometry in constant temperature mode (CTA) was employed to measure the velocity profile across the boundary layer at several streamwise positions, using a Dantec Dynamics StreamLine Pro system and an in-house repaired Dantec 55P05 boundary layer probe with a 5 $\mu$m diameter $d$ tungsten wire electroplated with copper. The wire had an exposed tungsten core approximately 1 mm in length $l$, achieving $l/d \geq 200$ to minimise end-conduction attenuation as recommended by \citet{Ligrani1987}. The overheat ratio was set to 0.8, corresponding to an over-temperature of approximately $220^\circ$C. Following square-wave testing, a 30 kHz low-pass filter was applied, and the sampling rate was set to $f_s = 60$ kHz, with a total acquisition time of 150 s per measurement point. This yielded a temporal resolution of $0.46 \leq t^+ \equiv U_\tau^2/(f_s\nu) \leq 0.74$ and satisfied statistical convergence based on boundary layer turnover times of $TU_{\infty}/\delta>20000$ as suggested by \citet{Hutchins2009}. The viscous-scaled spatial resolution of the probe ranged from $43 \leq l^+ \equiv lU_\tau/\nu \leq 53$ and remained below $l^+ \lesssim 60$, ensuring the validity of the analysis in the outer region of the TBL as suggested by \citet{Deshpande2023}, which is the primary region of interest in the present study. It should also be recognised that correction schemes for the attenuation of streamwise variance due to spatial averaging in non-canonical flows with APGs have been proposed \citep{Mallor2025}; however, as $l^+$ remains approximately constant across all cases in the present study, any attenuation effects are expected to be consistent, enabling meaningful relative comparisons without introducing systematic bias.

Hot-wire calibrations were performed before and after each measurement (approximately 4.5-hour intervals) using the \emph{in-situ} nearby Pitot tube (see \textcircled{5} in figure \ref{Figure:2}) measuring $U_{\infty,l}$ as the calibration reference. Both the hot-wire (see \textcircled{6} in figure \ref{Figure:2}) and pitot tube were mounted on an aerofoil-shaped traverse system with an encoder offering 1 $\mu$m resolution. During calibration, the wing (see \textcircled{4} in figure \ref{Figure:2}) angle was set to zero to avoid wake effects and ensure free-stream conditions. The calibration, performed following a King's law polynomial fit, allowed the hot-wire voltage output to be converted to velocity, with temperature correction applied as a precaution against potential temperature drift.

\subsection{\label{subsec:TBL properties}TBL properties and $U_{\tau}$ estimation}
The complete set of experimental conditions and key TBL characteristics relevant to the hot-wire measurements is summarised in table \ref{Table:2}. A key parameter in TBL studies is the skin-friction velocity, $U_\tau$, which characterises the skin-friction coefficient, $C_f$, and wall shear stress, $\tau_w$, according to $C_f = 2\tau_w/(\rho U_{\infty}^2) = 2(U_\tau/U_{\infty})^2$. In the present study, direct wall shear stress measurements were not conducted. Instead, $U_\tau$ was estimated using an indirect composite profile fitting method based on the measured boundary layer velocity profile. It has been shown that, for ZPG and mild APG TBLs (within the range of $\beta$ considered here, i.e. $0 \lesssim \beta \lesssim 2$), the Clauser chart method -- forming the basis of the present approach -- agrees well with  with direct oil-film interferometry measurements \citep{Monty2011}. The fitting procedure, following \citet{RodriguezLopez2015}, was used to determine $U_\tau$, the wall-normal coordinate correction $\upDelta y$, and the wake strength parameter $\mathit{\Pi}$ by collapsing the measured mean velocity onto a theoretical Musker profile combined with an exponential-type wake function. For all test cases listed in table \ref{Table:2}, the von K{\'a}rm{\'a}n constant was fixed at $\kappa = 0.39$ as discussed by \citet{Marusic2013}, with a corresponding log law intercept $B = 4.3386$ following the relation proposed by \citet{Nagib2008}. The indirect estimation of $U_\tau$ was applied to previously published datasets: PGH smooth-wall cases from \citet{Preskett2025a} and ZPG smooth-wall canonical cases from \citet{Wangsawijaya2023} and \citet{Aguiar2024}, which used direct measurements via oil-film interferometry and floating-element drag balance; the resulting differences were within 3\%, demonstrating the reliability of the method.

\begin{table}
\begin{center}
  \begin{tabular}{ccccccccccccccc}
  Case & Sym. & $x/\delta_{0}$ & $U_{\infty,l}$ & $\frac{\mathrm{d}C_p}{\mathrm{d}x}\delta_0$ & $U_\tau$ & $Re_\tau$ & $Re_\theta$ & $\delta$ & $\delta^{*}$ & $H$ & $\mathit{\Pi}$ & $\beta$ & $\frac{\mathrm{d}\beta}  {\mathrm{d}x} \delta_0$ & $\upDelta\beta$\\
        &       &      &   (ms$^{-1}$)   &  $(10^{-4})$  & (ms$^{-1}$) &      &      & (mm)  & (mm) &      &      &      &\\
  \hline
  \multicolumn{14}{c}{\textbf{Upstream of PGH}}\\
  $0^\circ$  &   & 41  & 20.16 & $-37$ & 0.80 & 1860 & 3645 & 35.23 & 3.99 & 1.46 & 0.26 & -0.13 & &\\
  \multicolumn{14}{c}{\textbf{Downstream of PGH}}\\
  $-4^\circ$ & \begin{tikzpicture}\node[circle, draw=b1, symbol style]{};\end{tikzpicture} & 109 & 21.52 & 134 & 0.69 & 2434 & 7158  & 53.34  & 8.47  & 1.67 & 1.33 & 1.33 & -0.15 & 0.88\\
                & \begin{tikzpicture}\node[circle, draw=b2, symbol style]{};\end{tikzpicture} & 122 & 21.61 & 24  & 0.71 & 3075 & 11 853 & 65.69  & 12.15 & 1.46 & 1.03 & 0.33 & -0.07 & 0.98\\
                & \begin{tikzpicture}\node[circle, draw=b3, symbol style]{};\end{tikzpicture} & 133 & 21.42 & 8   & 0.72 & 3295 & 10 433 & 69.64  & 11.32 & 1.53 & 0.88 & 0.09 & 0.02 & 0.68\\
                & \begin{tikzpicture}\node[circle, draw=b4, symbol style]{};\end{tikzpicture} & 147 & 21.51 & -4  & 0.73 & 3836 & 14 979 & 80.47  & 14.53 & 1.37 & 0.66 & -0.06 & 0.00 & 0.53\\
                & \begin{tikzpicture}\node[circle, draw=b5, symbol style]{};\end{tikzpicture} & 177 & 21.40 & -5  & 0.72 & 4807 & 18 510 & 102.33 & 17.78 & 1.34 & 0.67 & -0.10 & 0.00 & 0.16\\
                & \begin{tikzpicture}\node[circle, draw=b6, symbol style]{};\end{tikzpicture} & 191 & 21.69 & -1  & 0.74 & 5097 & 17 614 & 104.57 & 16.48 & 1.33 & 0.45 & -0.02 & 0.00 & 0.02\\
                & \begin{tikzpicture}\node[circle, draw=b7, symbol style]{};\end{tikzpicture} & 201 & 21.65 & -6  & 0.74 & 5319 & 18 095 & 108.89 & 16.83 & 1.33 & 0.45 & -0.10 & -0.01 & -0.01\\
  $-6^\circ$ & \begin{tikzpicture}\draw[g1, ultra thick, fill=none]
        (0,0.15) -- (0.12,0) -- (0,-0.15) -- (-0.12,0) -- cycle;\end{tikzpicture} & 109 & 21.88 & 161 & 0.68 & 2382 & 7961  & 53.65  & 9.47  & 1.71 & 1.59 & 1.88 & -0.15 & 1.28\\
                & \begin{tikzpicture}\draw[g2, ultra thick, fill=none]
        (0,0.15) -- (0.12,0) -- (0,-0.15) -- (-0.12,0) -- cycle;\end{tikzpicture} & 122 & 21.19 & 37  & 0.66 & 2952 & 12 427 & 67.32  & 13.55 & 1.53 & 1.35 & 0.63 & -0.11 & 1.52\\
                & \begin{tikzpicture}\draw[g3, ultra thick, fill=none]
        (0,0.15) -- (0.12,0) -- (0,-0.15) -- (-0.12,0) -- cycle;\end{tikzpicture} & 133 & 21.59 & 10  & 0.69 & 3354 & 12 757 & 73.48  & 13.68 & 1.52 & 1.13 & 0.17 & 0.00 & 1.11\\
                & \begin{tikzpicture}\draw[g4, ultra thick, fill=none]
        (0,0.15) -- (0.12,0) -- (0,-0.15) -- (-0.12,0) -- cycle;\end{tikzpicture} & 147 & 21.68 & -2   & 0.71 & 3986 & 16 542 & 84.71  & 16.34 & 1.41 & 0.94 & -0.04 & 0.00 & 0.82\\
                & \begin{tikzpicture}\draw[g5, ultra thick, fill=none]
        (0,0.15) -- (0.12,0) -- (0,-0.15) -- (-0.12,0) -- cycle;\end{tikzpicture} & 177 & 21.52 & -4  & 0.71 & 4999 & 19 967 & 107.96 & 19.36 & 1.36 & 0.80 & -0.09 & -0.01 & 0.32\\
                & \begin{tikzpicture}\draw[g6, ultra thick, fill=none]
        (0,0.15) -- (0.12,0) -- (0,-0.15) -- (-0.12,0) -- cycle;\end{tikzpicture} & 191 & 21.84 & 0  & 0.73 & 5535 & 20 226 & 114.99 & 19.00 & 1.34 & 0.64 & 0.00 & 0.00 & 0.12\\
                & \begin{tikzpicture}\draw[g7, ultra thick, fill=none]
        (0,0.15) -- (0.12,0) -- (0,-0.15) -- (-0.12,0) -- cycle;\end{tikzpicture} & 201 & 21.79 & -5  & 0.73 & 6049 & 22 015 & 124.41 & 20.39 & 1.34 & 0.47 & -0.10 & -0.01 & 0.08\\
  $-8^\circ$ & \begin{tikzpicture}\node[star, star points=5, star point ratio=2.5, draw=o1, symbol style]{};\end{tikzpicture} & 109 & 22.20 & 167 & 0.65 & 2313 & 7464  & 54.20  & 9.71  & 1.89 & 2.01 & 2.15 & -0.20 & 1.73\\
                & \begin{tikzpicture}\node[star, star points=5, star point ratio=2.5, draw=o2, symbol style]{};\end{tikzpicture} & 122 & 21.95 & 41  & 0.66 & 3083 & 14 442 & 70.59  & 15.38 & 1.55 & 1.57 & 0.80 & -0.13 & 1.92\\
                & \begin{tikzpicture}\node[star, star points=5, star point ratio=2.5, draw=o3, symbol style]{};\end{tikzpicture} & 133 & 21.98 & 14  & 0.67 & 3479 & 13 600 & 78.93  & 15.29 & 1.62 & 1.47 & 0.26 & 0.01 & 1.51\\
                & \begin{tikzpicture}\node[star, star points=5, star point ratio=2.5, draw=o4, symbol style]{};\end{tikzpicture} & 147 & 22.03 & 0   & 0.69 & 4260 & 18 419 & 93.24  & 18.60 & 1.47 & 1.12 & -0.01 & 0.00 & 1.23\\
                & \begin{tikzpicture}\node[star, star points=5, star point ratio=2.5, draw=o5, symbol style]{};\end{tikzpicture} & 177 & 21.75 & -4   & 0.69 & 5662 & 24 680 & 125.11 & 23.87 & 1.38 & 0.93 & -0.11 & -0.01 & 0.81\\
                & \begin{tikzpicture}\node[star, star points=5, star point ratio=2.5, draw=o6, symbol style]{};\end{tikzpicture} & 191 & 22.03 & 1  & 0.73 & 5959 & 23 248 & 125.52 & 21.74 & 1.35 & 0.60 & 0.03 & 0.00 & 0.32\\
  \end{tabular}
  \caption{Summary of flow parameters and key boundary layer properties. Reynolds number definitions are $Re_\tau \equiv U_\tau \delta / \nu$ and $Re_\theta \equiv U_{\infty,l} \theta / \nu$, where $\delta$ in this study is the boundary layer thickness based on $99\% U_{\infty,l}$, $\theta$ is the momentum thickness, and $\nu$ is the kinematic viscosity. The symbols $\delta^*$, $H$, $\mathit{\Pi}$, $\beta$, and $\upDelta\beta$ denote the displacement thickness, shape factor, wake strength parameter, Clauser PG parameter, and PGH integral parameter respectively.}
\label{Table:2}
  \end{center}
\end{table}

Regarding the PGH integral parameter, $\upDelta\beta$, it is observed that from the first to the second downstream location ($x/\delta_0 = 109$ to $122$), $\upDelta\beta$ increases by around 15\% despite a substantial reduction of around 65\% in $\beta$; at the same time, $\mathit{\Pi}$ decreases by approximately 20\%. Further downstream, $\mathit{\Pi}$ continues to recover towards the canonical ZPG value of $\mathit{\Pi} \approx 0.44$ \citep{Chauhan2009} (see table \ref{Table:2}). Indeed, once $\beta \approx 0$ and $\upDelta\beta \lesssim 0.1$, the corresponding $\Pi$ values are within approximately 5\% of the canonical ZPG value. This observation suggests that, although $\upDelta\beta$ is able to capture the history effect, $\beta$ may still need to be considered simultaneously to characterise the flow, e.g. for the prediction of $\upDelta\mathit{\Pi}$. It is also important to note that, as $\delta$ grows with downstream development, the integration length employed here, $L=30\delta$, increases continuously with downstream distance, ranging approximately from $45\delta_0$ to $100\delta_0$. Consequently, the effective upstream extent represented by $\upDelta\beta$ changes throughout the recovery process. For comparison, in \citet{Preskett2025a}, a length of $30\delta$ corresponded to approximately $45\delta_0$, comparable to the imposed impulsive PG length considered in their study. In the present work, however, $30\delta$ at $x/\delta_0=109$ in the weak PGH case corresponds to approximately half of the imposed PG length ($45\delta_0$), placing the upstream integration limit at $x_{\mathrm{upstream}} \approx 60\delta_0$; on the other hand, at $x/\delta_0=201$, the integration length increases to approximately the full length of the impulsive PG ($90\delta_0$), corresponding to $x_{\mathrm{upstream}} \approx 110\delta_0$. This suggests that further work towards the appropriate definition and physical interpretation of $L$ is required, potentially incorporating not only $\delta$ but also the characteristic lengthscale of the imposed PG itself. Such assessments, however, are beyond the scope of the present study.

\section{\label{sec:Matched Reynolds number}Matched-$Re_\tau$ comparison of recovering TBLs with varying PGH}
\begin{figure}
    \centering
    \includegraphics[width=\textwidth]{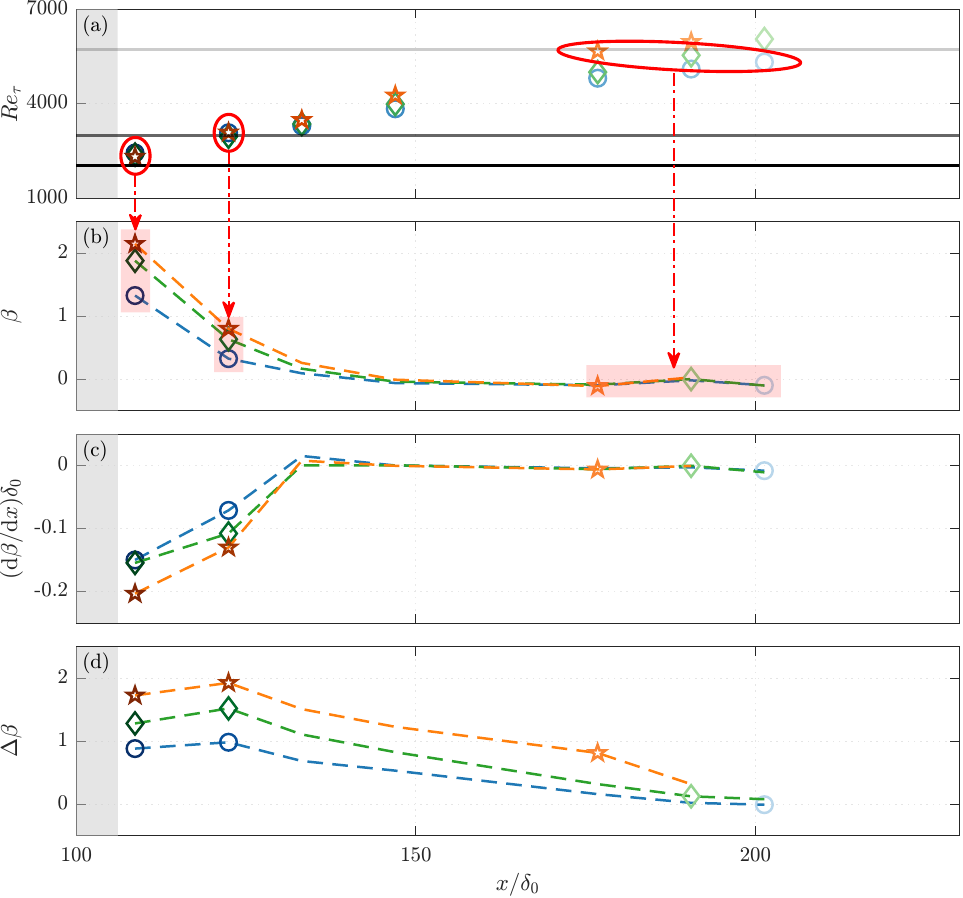}
    \caption{Grouping of experimental cases based on matched-$Re_\tau$ for all three PGH conditions. (a) Streamwise evolution of experimentally obtained $Re_\tau$ values. Red circles indicate selected cases grouped by similar $Re_\tau$, with corresponding baseline ZPG values from published datasets shown as horizontal lines. (b) Clauser PG parameter, $\beta$, for the grouped cases (shaded rectangles). (c) Streamwise gradient of the Clauser PG parameter, $\mathrm{d}\beta/\mathrm{d}x$, normalised by the incoming boundary-layer thickness $\delta_0$, highlighting the local disequilibration strength. (d) PGH integral parameter, $\upDelta\beta$. Symbols in (b)--(d) denote the matched-$Re_\tau$ grouped cases, while dashed lines are based on all other measurements, omitted for clarity. Full numerical values are provided in table \ref{Table:2}.}
    \label{Figure:4}
\end{figure}

To isolate the impact of PGH, the analysis is carried out between cases at matched-$Re_\tau$ as well as matched local $\beta$, allowing differences to be attributed primarily to PGH. Figure \ref{Figure:4}(a) shows all experimentally obtained $Re_\tau$ values, together with ZPG TBL data from previously published studies used for comparison.

The experimental cases are grouped by similar $Re_\tau$: measurements at $x/\delta_0 = 109$ are classified as $Re_\tau \approx 2300$, those at $x/\delta_0 = 122$ as $Re_\tau \approx 3000$, and those at $x/\delta_0 =$ 177, 191, and 201 -- corresponding from strongest to weakest PGH -- as $Re_\tau \approx 5500$. These matched Reynolds number groups are compared with ZPG datasets from previous studies at comparable Reynolds numbers. For the $Re_\tau \approx 2300$ group, the reference is the $Re_\tau = 2030$ dataset of \citet{Wangsawijaya2020}. For the $Re_\tau \approx 3000$ group, the $Re_\tau = 2998$ dataset of \citet{Medjnoun2026} is used for $y^+ \geq 50$, while for $y^+ < 50$ the $Re_\tau = 2030$ dataset of \citet{Wangsawijaya2020} is employed instead, owing to the unmatched spatial resolution $(l^+)$ in the former. For the $Re_\tau \approx 5500$ group, comparison is made with the $Re_\tau = 5710$ dataset, also from \citet{Medjnoun2026}.

The corresponding Clauser PG parameters, $\beta$, for each group are shown in figure \ref{Figure:4}(b), with $\beta = 1.74 \pm 0.41$, $0.57 \pm 0.24$, and $-0.06 \pm 0.06$ for the $Re_\tau \approx 2300$, $3000$, and $5500$ groups, respectively. Although some variation in $\beta$ exists among the matched-$Re_\tau$ groups between the weak, mild, and strong PGH cases, these differences remain secondary compared to the substantially larger divergence in imposed PGH amplitudes (see \S~\ref{subsec:Flow history}). The corresponding normalised local disequilibrium parameter, $(\mathrm{d}\beta/\mathrm{d}x)\delta_0$, is shown in figure \ref{Figure:4}(c). Note that $\mathrm{d}\beta/\mathrm{d}x$ should be interpreted as an estimate, since $\mathrm{d}\delta^*/\mathrm{d}x$ and $\mathrm{d}\tau_w/\mathrm{d}x$ are obtained using linear interpolation from limited measurement stations. Nevertheless, the estimation remains primarily governed by the pressure-term contribution $\mathrm{d}^2p/\mathrm{d}x^2$ obtained from the more spatially resolved pressure taps measurements (see \S~\ref{subsec:Flow history}), rather than from direct differentiation of the $\beta(x)$ curve. At the first downstream location, where the largest absolute values occur in the present dataset, the equivalent local disequilibrium parameters $\mathrm{d}\beta/\mathrm{d}Re_\tau$ and $\mathrm{d}\beta/\mathrm{d}Re_\theta$ are of order \textit{O}$(10^{-3})$ and \textit{O}$(10^{-4})$, respectively, with variations between PGH cases within 10\%. These values are comparable to those analysed by \citet{Mahajan2026} as near-equilibrium TBLs. Although the first two matched-$Re_\tau$ groups are not strictly free from local disequilibrium effects, the relatively small and comparable values within each matched-$Re_\tau$ group suggest that the dominant contribution to the observed behaviour arises from PGH effects.

The corresponding PGH integral parameter, $\upDelta\beta$, shown in figure \ref{Figure:4}(d), exhibits clear and systematic differences across the weak, mild, and strong PGH cases, spanning approximately $\upDelta \beta \approx 0-2$ across all groups. Most notably, in the $Re_\tau \approx 5500$ group, despite the flow having relaxed to nominally ZPG conditions ($\beta \approx 0$) and having remained so for more than approximately $30\delta_0$, $\upDelta\beta$ remains markedly different between cases, with values of -0.01, 0.12, and 0.81 for the weak, mild, and strong cases, respectively. $30\delta_0$ is indeed a long streamwise distance; relative to the local $\delta$ of the strong PGH case, this is equivalent to more than approximately $8\delta$, exceeding the distance required to reach near-equilibrium of $7\delta$ reported by \citet{Bobke2017}. This indicates that $\upDelta\beta$ effectively captures the cumulative effect of the upstream history, such that any observed discrepancies in the flow at this downstream location can be attributed entirely to history effects. This directly supports the rationale for grouping the data by $Re_\tau$ and isolating PGH as the primary source of variation.

\subsection{\label{subsec:MVP and TI}Mean velocity and turbulence intensity}
\begin{figure}
    \centering
    \includegraphics[width=\textwidth]{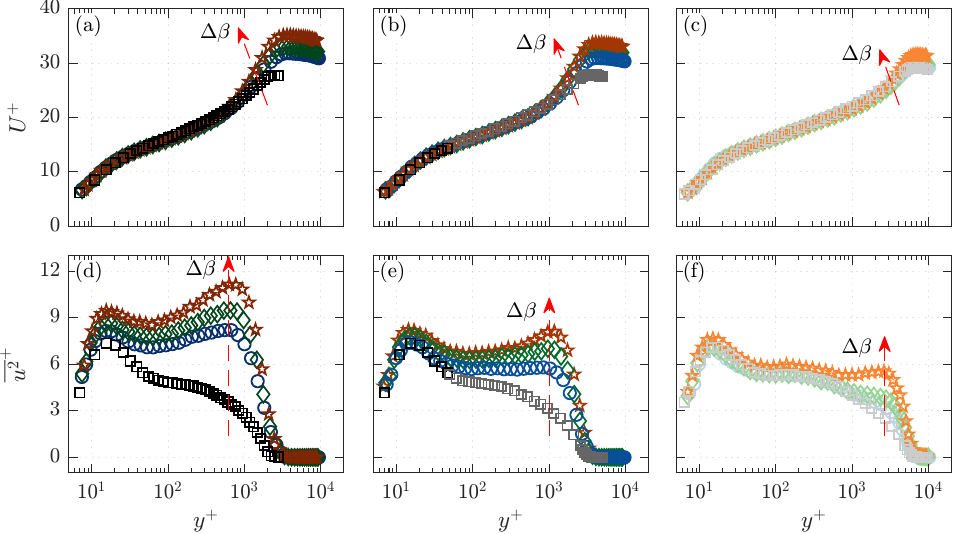}
    \caption{Inner-scaled mean velocity profiles for matched friction Reynolds number of  $Re_\tau\approx$ (a) 2300, (b) 3000, and (c) 5500, with the corresponding inner-scaled turbulence intensity profiles shown in (d), (e), and (f). \protect\begin{tikzpicture}
        \protect\node[rectangle, draw=k1, symbol style]{};
    \protect\end{tikzpicture}, 
    \protect\begin{tikzpicture}
        \protect\node[rectangle, draw=k2, symbol style]{};
    \protect\end{tikzpicture}, and
    \protect\begin{tikzpicture}
        \protect\node[rectangle, draw=k3, symbol style]{};
    \protect\end{tikzpicture}
    denote the reference ZPG data at $Re_\tau=$ 2030, 2998, and 5710, respectively. $\upDelta \beta$ serves as a measure of PGH strength. The coloured symbols correspond to the test cases listed in table \ref{Table:2}.}
    \label{Figure:5}
\end{figure}

In figure \ref{Figure:5}, the TBL statistics of the different PGH cases are presented for all matched-$Re_\tau$ groups, compared with the baseline ZPG reference case. Panels (a), (b), and (c) show the viscous-scaled mean streamwise velocity profiles $(U^+)$
plotted against the viscous-scaled wall-normal coordinate $y^+$ for $Re_\tau \approx 2300$, $3000$, and $5500$, respectively. Measurements extending into the lower buffer layer $(y^+ \lesssim 8)$ provide detailed insights into the near-wall flow behaviour.

Just downstream of the wing, at $Re_\tau \approx 2300$ where PGH influence is expected to be strong, the mean velocity profiles show prominent differences in the wake region. The stronger PGH case exhibits a more pronounced wake -- even the weak case displays a significantly stronger wake than the ZPG reference -- evidenced by higher $U^+$ in the freestream and larger $\mathit{\Pi}$ values in table \ref{Table:2}. The intensified wake is accompanied by a shortening of the logarithmic region due to the earlier onset of wake effects. These behaviours are consistent with the established understanding of APG impacts on TBLs, as reported by \citet{Harun2013} and \citet{Vinuesa2017b}. The inner region, however, shows excellent collapse across all PGH cases and the ZPG reference, reinforcing the classical observation by \citet{Clauser1956} that the inner layer remains largely insensitive to PGs, provided flow separation does not occur.

Further downstream, as the flow reaches $Re_\tau \approx 3000$, the profiles retain the same overall behaviour: increasingly pronounced wake regions with stronger PGH, and excellent collapse in the inner layer across all cases. Compared to the previous station, the wake strength decreases and the logarithmic region extends. While this extension may partly result from the increasing $Re_\tau$, the concurrent reduction in $\mathit{\Pi}$ (see table \ref{Table:2}) supports the interpretation of a diminishing APG influence.

At $Re_\tau \approx 5500$, the local PG has fully diminished, with $\beta$ having relaxed to values nominally zero. The weak PGH case collapses well with the ZPG data, whereas the stronger PGH cases continue to exhibit enhanced wake strength, most pronounced in the strongest case. This is also reflected in the numerical values of $\mathit{\Pi}$ in table \ref{Table:2}, which remain elevated for the mild and strong cases but recover towards the canonical ZPG value for the weak case. It is worth noting that the weak PGH case corresponds to $\upDelta\beta \approx 0$, and consequently exhibits a good collapse with the ZPG reference. By contrast, the mild and strong PGH cases retain non-zero $\upDelta\beta$, and accordingly display a persistent deviation in $\mathit{\Pi}$ from the canonical values. This demonstrates the ability of $\upDelta\beta$ to capture the cumulative effect of PGH on the mean flow. Additionally, we observed that the shape factor $H$ (see table \ref{Table:2}), which increases under APG, recovers much earlier -- at the fourth downstream location -- to its typical ZPG range of 1.3 to 1.4, consistent with \citet{Schlichting2018}, once $\mathrm{d}C_p/\mathrm{d}x$ has relaxed. These observation suggest that while $H$ primarily reflects the influence of local $\mathrm{d}C_p/\mathrm{d}x$, $\mathit{\Pi}$ retains a stronger imprint of PGH which is captured by $\upDelta\beta$. It is evident that the flow preserves memory of its upstream history, even once $\beta$ has relaxed. Nevertheless, the gradual convergence of mean flow properties towards those of a canonical ZPG TBL indicates the existence of a recovery process from PGH effects.

The corresponding inner-scaled streamwise Reynolds stress $(\overline{u^2}^+ \equiv \overline{u^2}/U_\tau^2)$ 
are shown in figure \ref{Figure:5}(d), (e), and (f) for $Re_\tau \approx 2300$, $3000$, and $5500$, respectively. For $Re_\tau \approx 2300$, near-wall measurements reveal a clear inner peak, whose magnitude increases with PGH strength, although its location remains approximately fixed at $y^+ \approx 15$. The turbulence intensity throughout the logarithmic region is also elevated, increasing with PGH strength. At $y^+ \approx 600 \approx 0.26 Re_\tau$, an outer peak appears in all the PGH cases, whereas it is absent in the ZPG case, with stronger PGH yielding higher magnitudes. This outer peak is a well-known signature of APG effects and should not be confused with the VLSM-induced hump typically observed in very high-$Re_\tau$ TBLs near the centre of the logarithmic region \citep{Harun2013}. These results indicate that, while the mean flow near the wall shows little footprint of PGH, the turbulence variance reveals otherwise, with APG amplifying fluctuations across all regions of the TBL.

At $Re_\tau \approx 3000$, amplification of the inner peak remains evident, but in the weak PGH case it has recovered and coincides with the ZPG reference up to the end of the buffer layer at $y^+ \approx 30$. In the outer region, amplification across the layer remains evident, but it is weaker compared with the $Re_\tau \approx 2300$ cases upstream. The location of these features appears insensitive to PGH strength, remaining at $y^+ \approx 990$, or approximately $y/\delta \approx 0.33$. Both the mild and strong PGH cases still show elevated Reynolds stress throughout the log region, although now appearing as a broad energetic plateau. Interestingly, even the weak PGH case -- despite its inner peak returning to a ZPG-like state -- continues to exhibit enhanced energy across the log region.

Further downstream, at $Re_\tau \approx 5500$, the amplification of turbulence intensity is weaker. For the weak and mild PGH cases, both the inner peak and logarithmic region collapse with the ZPG reference, whereas the strong PGH case continues to exhibit elevated turbulence across the boundary layer. Despite this recovery, the outer peak persists: it appears as a subtle hump in the weak and mild cases, and remains clearly amplified in the strong case, with an additional peak emerging at $y^+ \approx 140$ within the log region. Taken together, these results suggest that the modification of the logarithmic layer is driven by the persistence of outer-layer structures associated with the outer peak, which maintain a lasting footprint of PGH even after the mean flow recovers to ZPG conditions. To further investigate which turbulent scales contribute to these PGH-induced modifications, an energy spectral analysis of the streamwise velocity fluctuations is presented in the following section.

\subsection{\label{subsec:Spectra}Energy spectra}
\begin{figure}
    \centering
    \includegraphics[width=\textwidth]{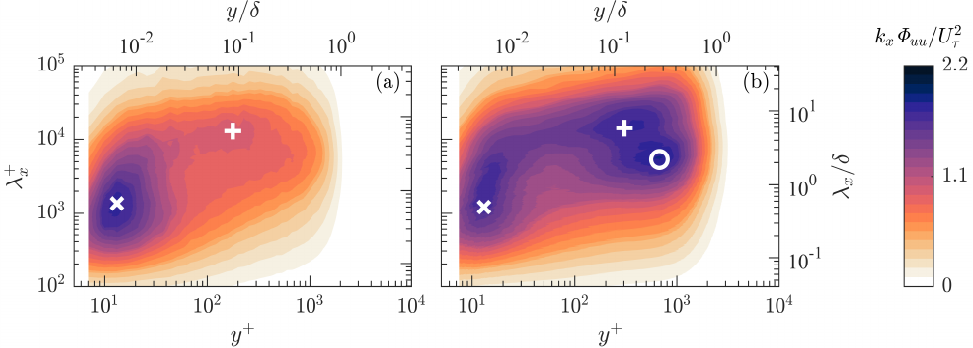}
    \caption{Premultiplied one-dimensional energy spectra of streamwise velocity fluctuations at $Re_\tau \approx 2300$ for (a) the ZPG case and (b) the weak PGH case ($-4^\circ$). Both cases exhibit the inner-site (`$\times$') and VLSM (`$+$') peaks, while the weak PGH case additionally shows a PG peak (`$\circ$'), highlighting the effect of PGH on the spectra.}
    \label{Figure:6}
\end{figure}

Figure \ref{Figure:6} shows the inner-scaled one-dimensional premultiplied spectra, $k_x \mathit{\Phi}_{uu}/U_\tau^2$ (where $k_x$ is the streamwise wave number and $\mathit{\Phi}_{uu}$ is the power spectral density of the streamwise velocity fluctuations), plotted against $y^+$ and inner-scaled streamwise wavelength $\lambda_x^+ \equiv \lambda_x U_\tau/\nu$, comparing (a) the ZPG reference case at $Re_\tau = 2030$ and (b) the weak PGH case at $Re_\tau = 2434$ and $\beta = 1.33$.

In the ZPG reference (figure \ref{Figure:6}a), two energetic features are evident. The first is a small-scale near-wall peak at $\lambda_x^+ \approx 1000$ and $y^+ \approx 15$, corresponding to the inner-site peak described by \citet{Hutchins2007} and associated with the energetic contribution from the near-wall cycle. The second is a very-large-scale outer-site energetic content at $\lambda_x \approx 6\delta$ and $y \approx 0.06\delta$, consistent with the VLSM peak reported by \citet{Mathis2009}. 

With weak PGH case (figure \ref{Figure:6}b), \emph{inner-site peak} remains at $\lambda_x^+ \approx 1000$ and $y^+ \approx 15$, with increased magnitude, as also visible in the variance plot (figure \ref{Figure:5}d). In the log and outer regions, energy is amplified consistent with APG effects noted by \citet{Monty2011} and also apparent in the variance plot (figure \ref{Figure:5}d). A VLSM peak is observed at $y \approx 0.12\delta$, $\lambda_x \approx 6\delta$, while an additional, smaller-scale outer peak appears in the wake region at $y \approx 0.3\delta$, $\lambda_x \approx 2.2\delta$. In contrast with \citet{Harun2013} and \citet{Deshpande2023}, who reported a single broad outer enhancement under APG, the present results reveal two distinct outer peaks, here denoted as the \emph{VLSM peak} (`$+$') and the \emph{PG peak} (`$\circ$') in figure \ref{Figure:6}(b).

\begin{figure}
    \centering
    \includegraphics[width=\textwidth]{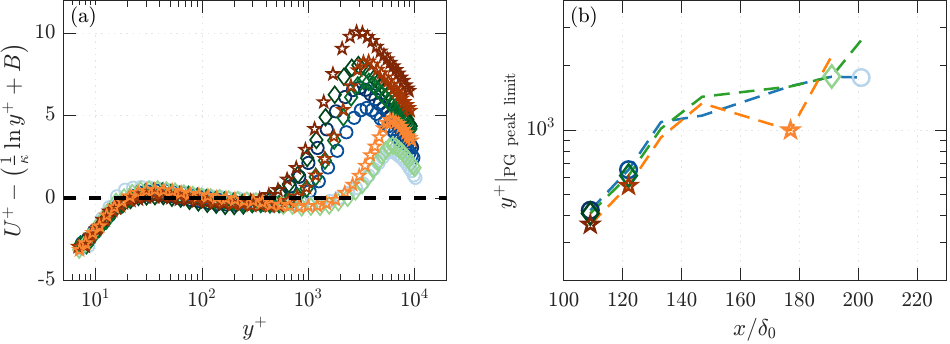}
    \caption{Estimation of the threshold wall-normal position $y^+|_\mathrm{PG\ peak\ limit}$ used to distinguish PG from VLSM peaks. Panel (a) shows the mean velocity profile $U^+$ subtracted by the log law $(1/\kappa)\ln y^+ + B$ for all downstream locations, with the horizontal dashed line indicates the log law. Panel (b) presents the corresponding wall-normal positions $y^+|_{\mathrm{PG\ peak\ limit}}$ -- dashed lines represent other cases, with symbols omitted for clarity -- separating the PG peak region from the VLSM zone.}
    \label{Figure:7}
\end{figure}

Before examining the spectra of the other PGH cases, we first seek to isolate the PG peak from the VLSM peak. To aid this distinction, we define the edge of the logarithmic region (where the VLSM peak is located) and the onset of the wake region (where the PG peak arises). The diagnostic plot, previously used by \citet{Marusic2013} to assess the consistency of logarithmic behaviour with a universal von K{\'a}rm{\'a}n constant, enables identification of the wake onset due to APG effects, as shown in figure \ref{Figure:7}(a).

Accordingly, we define the threshold wall-normal location $y^+|_{\mathrm{PG\ peak\ limit}}$ as the point marking the demarcation from the log region into the onset of the wake region, where deviation from the logarithmic profile becomes significant:
\begin{equation}
    \frac{\mathrm{d}\left\langle U^+ - \left(\frac{1}{\kappa} \ln y^+ + B\right) \right\rangle}{\mathrm{d}\left\langle \ln y^+ \right\rangle} \geq 1, \quad y^+ \geq 30.
\end{equation} 
This threshold is illustrated in figure \ref{Figure:7}(b) and shown by vertical dashed lines in all spectra plots of the PGH cases. With this criterion, the two outer peaks become more discernible: the peak below $y^+|_{\mathrm{PG\ peak\ limit}}$ is classified as the VLSM peak, while the one above it is identified as the PG peak.

\begin{figure}
    \centering
    \rotatebox{90}{
        \begin{minipage}{\textheight}
            \centering
            \includegraphics[width=0.95\textheight]{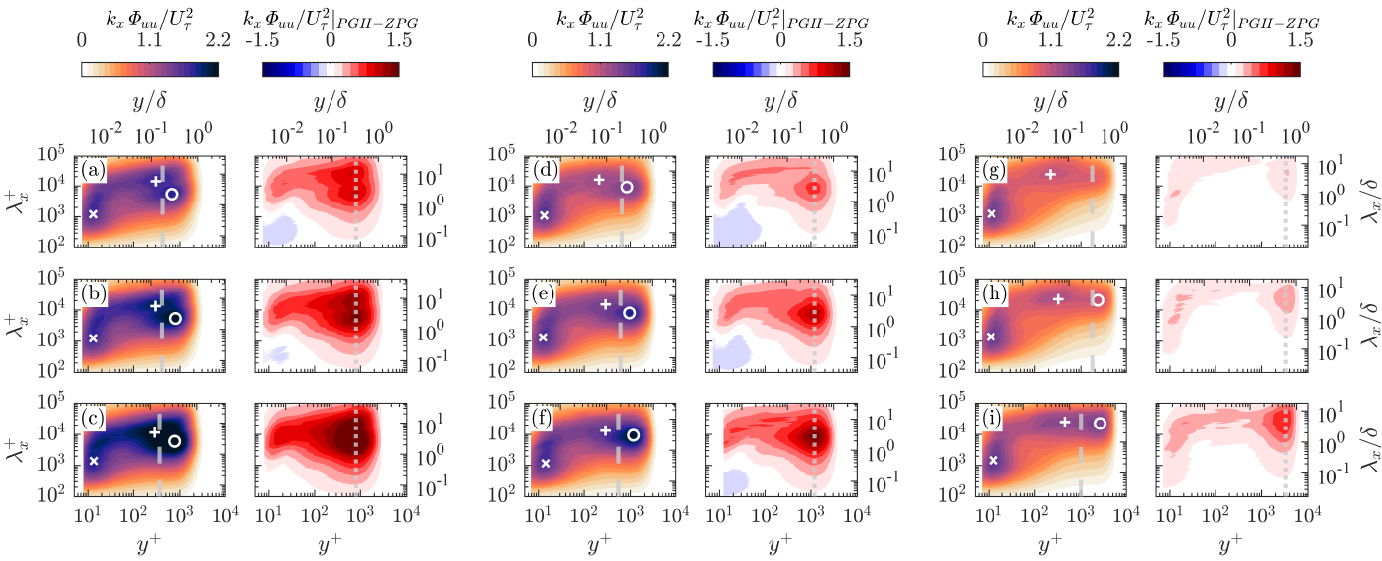}
            \captionsetup{width=0.95\textheight}
            \caption{One-dimensional premultiplied streamwise velocity spectra for matched-$Re_\tau$ comparisons across PGH cases. Subfigures (a--c) correspond to $Re_\tau \approx 2300$, (d--f) to $Re_\tau \approx 3000$, and (g--i) to $Re_\tau \approx 5500$, with each set representing PGH cases of $-4^\circ$, $-6^\circ$, and $-8^\circ$, respectively. In each subfigure, the left panel shows the premultiplied spectra $k_x\mathit{\Phi}_{uu}/U_\tau^2$, and the right panel shows the difference relative to the matched-$Re_\tau$ ZPG TBL, obtained by subtracting the ZPG spectrum from the PGH case. In the spectral maps, the symbols `$\times$', `$+$', and `$\circ$' denote the inner-site, VLSM, and PG peaks, respectively. The vertical dashed line in the left panel marks the threshold $y^+|_\mathrm{PG\ peak\ limit}$, while the vertical dotted line in the right panel indicates the core energy alteration location used to delineate the PG peak region.}
            \label{Figure:8}
        \end{minipage}
    }
\end{figure}

To further investigate the structural imprints of PGH on the energy distribution across scales, figure \ref{Figure:8} presents the spectra for each PGH case at matched-$Re_\tau$, providing a focused view of how PGH strength influences the redistribution of energy across the boundary layer and turbulent scales. The left panels in each subfigure show the inner-scaled premultiplied spectra, while the right panels display the difference from the corresponding ZPG reference, thereby highlighting the net spectral modification introduced by PGH.

The spectra in Fig.\ref{Figure:8}(a) correspond to the weak PGH case at $Re_\tau \approx 2300$, showing the three peaks already discussed in figure \ref{Figure:6}(b). Here, the vertical dashed lines mark the threshold separating the VLSM peak (left of the line) from the PG peak (right of the line). As the PGH strength increases to mild and strong cases at $Re_\tau \approx 2300$ (figure \ref{Figure:8}(b--c)), the three characteristic peaks remain at approximately $y^+ \approx 15$ and $\lambda_x^+ \approx 1000$ for the inner-site, $y/\delta \approx 0.12$ and $\lambda_x/\delta \approx 6$ for the VLSM peak, and $y/\delta \approx 0.3$ and $\lambda_x/\delta \approx 2.2$ for the PG peak. While their wall-normal locations and scales remain essentially unchanged, the energy intensity of all three peaks increases with stronger PGH, consistent with the variance trends in figure \ref{Figure:5}(d). The spectra also reveal a small-scale energy deficit accompanying the amplification of large-scale energy in the inner-site, similar to that reported by \citet{Harun2013}. It is also worth noting that, despite the strong energy amplification in the outer region for the strong PGH case, the inner-site peak remains clearly identifiable, suggesting that near-wall streaks persist as a turbulence regeneration mechanism even under the dominance of outer-layer large-scale motions, echoing the interpretation of \citet{Gungor2022}.

At the second downstream location, $Re_\tau \approx 3000$ (figure \ref{Figure:8}(d--f)), three distinct peaks persist across all PGH strengths. The inner-site peak (`$\times$' in figure \ref{Figure:8}) remains essentially unchanged compared to the previous downstream position. The VLSM peak (`$+$' in figure \ref{Figure:8}) shifts closer to its typical position while occuring at smaller streamwise scales, now appearing at $y/\delta \approx 0.09$ and $\lambda_x/\delta \approx 4.7$. In contrast, the PG peak (`$\circ$' in figure \ref{Figure:8}) moves further from the wall and occurs at larger streamwise scales, now located at $y/\delta \approx 0.32$ and $\lambda_x/\delta \approx 2.9$. Examination of the spectral difference maps for the PG peak region shows that the core enhancement remains centred at $y/\delta \approx 0.4$ (indicated by the vertical dotted line) and $\lambda_x/\delta \approx 2.5$ across all cases, while the magnitude of energy amplification increases with PGH strength.

Further downstream, at $Re_\tau \approx 5500$ (figure \ref{Figure:8}(g--i)), the spectral peak features remain broadly similar. In the weak PGH case (figure \ref{Figure:8}g), the PG peak has vanished, leaving only the inner-site and VLSM peaks, which resemble the canonical ZPG state. By contrast, in the mild and strong PGH cases (figure \ref{Figure:8}h,i), all three peaks persist. The inner-site peak remains essentially unchanged, while the VLSM peak has returned to its canonical position but occurs at a shorter streamwise wavelength than upstream ($y/\delta \approx 0.06$, $\lambda_x/\delta \approx 4.5$). The PG peak, present only in the mild and strong cases, has moved further from the wall and occurs at larger streamwise scales ($y/\delta \approx 0.4$, $\lambda_x/\delta \approx 4$). Notably, even though the PG peak is absent in the weak case, the spectral difference maps still reveal a residual core enhancement at $y/\delta \approx 0.4$ (see the dashed vertical line in the right panels of figure \ref{Figure:8}g-i) and $\lambda_x/\delta \approx 4$, highlighting the lasting impact of PGH.

Across the matched-$Re_\tau$ spectra analysis, PGH strength primarily affects the energy intensity of the peaks -- especially in the outer-layer -- while their spatial characteristics (wall-normal position and streamwise scale) remain largely invariant at a given $Re_\tau$. With increasing downstream distance from the wing, the locations and scales of the VLSM and PG peaks progressively reorganise, suggesting that PGH plays a central role in the redistribution of large-scale structures and the eventual recovery towards ZPG conditions.

\section{\label{sec:Recovery observation}Recovery observation: evolution of turbulent scales}
\begin{figure}
    \centering
    \includegraphics[width=\textwidth]{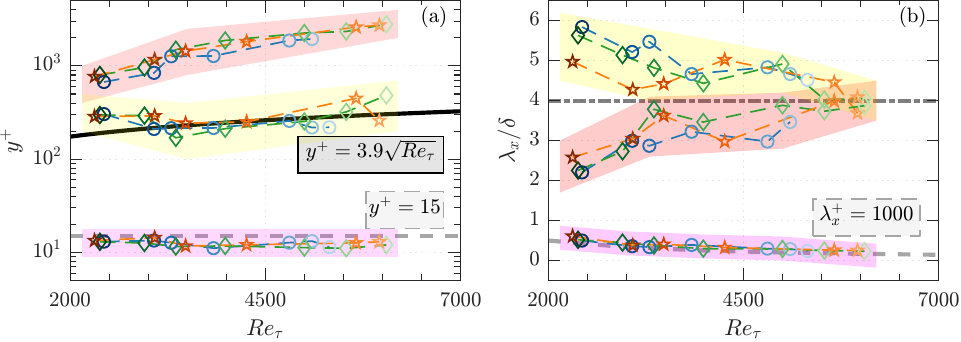}
    \caption{Downstream evolution of the wall-normal position and streamwise wavelength of the identified spectral peaks. Panel (a) shows the inner-scaled wall-normal location, $y^+$, while panel (b) shows the corresponding outer-scaled wavelength, $\lambda_x / \delta$, of the peaks. Each panel traces the evolution of the three key spectral features: inner-site peak (purple), VLSM peak (yellow), and PG peak (red). In panel (a), the curve denotes $y^+ = 3.9\sqrt{Re_\tau}$, indicating the geometric centre of the logarithmic region where the VLSM peak typically resides \citep{Mathis2009}, while the horizontal dashed line at $y^+ = 15$ marks the typical location of the inner-site peak \citep{Hutchins2007}. In panel (b), the dashed curve denotes $\lambda_x^+ = 1000$, the typical streamwise length of the inner-site peak \citep{Hutchins2007}, while the dash-dotted line marks $\lambda_x / \delta = 4$, associated with large-scale merging due to history effects.}
    \label{Figure:9}
\end{figure}

To further investigate the reorganisation of turbulent structures during recovery towards a canonical state, figure \ref{Figure:9} presents the downstream evolution of the spectral peak locations and associated length scales across all cases listed in table \ref{Table:2}, rather than restricting the analysis to matched $Re_\tau$ and $\beta$ conditions.

The inner-site peak remains fixed at $y^+ \approx 15$ and $\lambda_x^+ \approx 1000$ (purple highlight in figure \ref{Figure:9}), consistent with \citet{Hutchins2007}, and is largely unaffected by PGH or $Re_\tau$. This invariance contrasts with the behaviour of the outer-site peaks, namely the VLSM and PG peaks.

For the wall-normal position (figure \ref{Figure:9}a), the VLSM peak (yellow highlight) deviates from the canonical empirical scaling of the logarithmic region centre, $y^+ \approx 3.9\sqrt{Re_\tau}$, corresponding to $y/\delta \approx 0.06$, at the first two downstream locations ($Re_\tau \lesssim 3000$), where the local APG displaces it outward. Beyond this point, once the local PG has relaxed to zero, the VLSM peak recovers to its canonical position -- indicating that its wall-normal location is governed primarily by $Re_\tau$ and the local flow condition. On the other hand, the PG peak (red highlight) remains located further from the wall than the VLSM and continues to shift outward during recovery, generally residing within $0.3 \lesssim y/\delta \lesssim 0.5$.

Regarding the streamwise wavelength (figure \ref{Figure:9}b), the VLSM peak (yellow highlight) at the first downstream position appears at $\lambda_x / \delta \approx 6$, similar to the canonical value \citep{Hutchins2007} -- though slightly reduced to $\lambda_x / \delta \approx 5$ for the strong PGH case. As the flow develops, its scale progressively shortens toward $\lambda_x / \delta \approx 4$, a trend that continues even once the local $\beta$ has relaxed to zero. By contrast, the PG peak (red highlight) at $Re_\tau \approx 2300$ occurs at $\lambda_x$ between 2--3$\delta$, within the range of typical LSMs and consistent with the APG scale range reported by \citet{Harun2013}. As flow develops and the local $\beta$ relaxes to zero, this peak gradually lengthens toward $\lambda_x / \delta \approx 4$. The convergence of the VLSM and PG peak length scales toward $\lambda_x \approx 4\delta$ (marked by the horizontal dash-dotted line in figure \ref{Figure:9}b) indicates a recovery process in which \emph{history} effects play a more significant role than the \emph{local} PG.

Combining the analysis of outer-region energy spectra from figures \ref{Figure:8} and \ref{Figure:9}, it can be concluded that PGH strength primarily governs the energy amplification (as also observed in APG for constant $\beta$ \citep{Monty2011, Bobke2017, Deshpande2023}), while the \emph{history} (the combination of distance from the wing and $Re_\tau$) dictates the spatial distribution of turbulent scales. Notably, in the weak PGH case at $Re_\tau \approx 5500$ (figure \ref{Figure:8}g), although the spectra resemble the canonical ZPG state with no PG peak (only inner-site and VLSM peaks), figure \ref{Figure:9} shows that the VLSM peak, while returning to its typical wall-normal position determined by $Re_\tau$, exhibits a reduced streamwise lengthscale by roughly $2\delta$ -- indicating the lasting influence of PGH on turbulence structure. A possible explanation of such observations is reminiscent of the interpretation of \citet{Alving1990} -- although their study considered a different type of perturbation -- who suggested that large-scale structures may reorganise into a persistent, new quasi-stable state, effectively retaining their identity and leading to unusually long recovery distance in the outer-layer.

\section{\label{sec:Conclusion}Conclusion}
This study experimentally examined the \emph{recovery} of TBLs from PGH effects further downstream after the local PG \emph{relaxes}, using hot-wire anemometry downstream of a spatially impulsive FAPG sequence to assess the influence of history on flow behaviour. Comparative analysis was carried out with data at $Re_\tau \approx 2300$, $3000$, and $5500$ where the local $\beta$ was approximately matched at $1.74$, $0.57$, and $-0.06$, respectively. The $Re_\tau \approx 2300$ group corresponds to the location closest to the impulsive PG, whereas $Re_\tau \approx 5500$ corresponds to the furthest downstream location. At each of these matched $Re_\tau$ and $\beta$ groups, the TBL was subjected to three PGH strengths -- weak, mild, and strong -- characterised by $\upDelta\beta$ and the results were compared against canonical ZPG TBL datasets. Some of the key findings are:

(i) The mean velocity profiles, $U^+$, show that PGH strongly affects the wake region, with stronger PGH leading to a more pronounced wake, reflected in higher values of $\mathit{\Pi}$. In contrast, the inner layer collapses well across all cases, indicating its insensitivity to PGH. A recovered mean flow is observed once $\upDelta\beta \lesssim 0.1$, with $\mathit{\Pi}$ closely matching the canonical value. This is evident at the $Re_\tau \approx 5500$ group, where the weak PGH case collapses with the ZPG reference, and $\mathit{\Pi}$ lies within 5\% of the canonical value. By contrast, although $\beta$ has relaxed to nominal ZPG conditions for more than $30\delta_0$, the mild and strong PGH cases retain an enhanced wake, as captured by the non-zero $\upDelta\beta$.

(ii) The streamwise Reynolds stress profiles, $\overline{u^2}^+$, exhibit broadband modifications characterised by an enhanced inner peak, elevated logarithmic-layer energy, and the emergence of an outer peak, all of which intensify with increasing PGH strength. In the $Re_\tau \approx 5500$ group, despite $\beta$ having relaxed to zero, the strong PGH case retains elevated $\overline{u^2}^+$ throughout the boundary layer. Notably, even in the weak PGH case -- where the mean flow is effectively recovered -- the outer-layer Reynolds stress remains altered despite recovery of the inner peak and logarithmic region.

(iii) Spectral analysis shows that PGH primarily modifies outer-layer energy, with PGH strength governing the degree of energy amplification. The inner-site peak remains clearly identifiable and largely unaffected across all cases, indicating robustness of near-wall small-scale dynamics even under strong large-scale energetic activity. In the outer-site, the VLSM peak is energised by PGH but remains distinguishable, while an additional outer-layer feature emerges at smaller streamwise scales than the VLSM, identified here as the \emph{PG peak}, with $\lambda_x \approx 2$--$3\delta$. At $Re_\tau \approx 5500$, the PG peak is no longer discernible in the weak PGH case but persists in the mild and strong cases. Nevertheless, residual outer-layer energy enhancement remains even when the PG peak is absent, indicating a lasting imprint of PGH on the turbulence structures. This persistent footprint explains the observed deviations in streamwise Reynolds stress despite recovery of the mean flow.

Furthermore, regarding the evolution of the energetic scales during recovery, we find that:

(i) The wall-normal position of the VLSM peak returns towards its canonical location during downstream recovery, governed by $y^+ \approx 3.9\sqrt{Re_\tau}$ (corresponding to $y/\delta \approx 0.06$). In contrast, the PG peak remains further from the wall and continues to migrate outward during recovery, residing within $0.3 \lesssim y/\delta \lesssim 0.5$.

(ii) Although the VLSM peak recovers its canonical wall-normal position, its streamwise lengthscale progressively shortens from $\lambda_x \approx 6\delta$ to $\lambda_x \approx 4\delta$. Simultaneously, the PG peak lengthens from approximately $2$--$3\delta$ towards $\lambda_x \approx 4\delta$, where it eventually merges with the `recovered' VLSM peak. The term `recovered' is used here with caution, since the turbulence characteristics -- in this case the VLSM peak -- do not fully recover to the canonical state. We observe that the shortening of the VLSM lengthscale by approximately $2\delta$ persists even after the PG peak has vanished, indicating that reorganisation of the large-scale structures may underlie the long recovery distance -- emphasising that \emph{history} matters. The present data cannot determine whether or when full recovery (i.e. complete agreement of the turbulence characteristics with the canonical state) occurs, as measurements further downstream are not available.

Finally, we report that although $\upDelta\beta$ effectively captures deviations of $\mathit{\Pi}$ from the canonical state due to PGH effects, the local value of $\beta$ may still need to be considered simultaneously to characterise the flow. Moreover, further work towards a clearer interpretation of the integration length $L$ is needed, which may require consideration not only of $\delta$ but also of the characteristic lengthscale of the imposed PG (i.e. the streamwise extent over which the TBL is exposed to the PG relative to $\delta_0$). Future studies should explore these points, not only to confirm the observed evolution of turbulent scales during recovery, but also to better understand the persistent role of \emph{history} in TBL evolution.

\begin{bmhead}[Funding.]
We gratefully acknowledge the financial support from EPSRC (Grant Ref no: EP/W026090/1). DDW acknowledges the financial support from Leverhulme Trust (Grant no. ECF-2022-295).
\end{bmhead}
\begin{bmhead}[Declaration of interests.]
The authors report no conflict of interest.
\end{bmhead}
\begin{bmhead}[Data availability statement.]
The data that support the findings of this study will be made available upon publication at the University of Southampton's repository.
\end{bmhead}
\begin{bmhead}[Author contributions.]
ZB designed the setup, conducted experiments, and post-processed the hot-wire data as well wrote the first draft. DDW supervised ZB and edited drafts. BG conceived the work, acquired funding and edited drafts.
\end{bmhead}

\bibliographystyle{jfm}
\bibliography{jfm}

\end{document}